# Revolutionizing Mobility: The Latest Advancements in Autonomous Vehicle Technology

**Venkata Sai Chandra Prasanth Narisett[1]y,**
**Tejaswi Maddineni[2]**

[1]*Quality Engineer, Independent Researcher, Southern Illinois University Carbondale, United States*
[2]*Data Engineer, Independent Researcher, Southern Illinois University Carbondale, United States*
E-mail: [1]venkatn0388@gmail.com,
[2]tejas.maddineni@gmail.com.

**Abstract**

Autonomous vehicle (AV) technology is transforming the landscape of transportation by promising safer, more efficient, and sustainable mobility solutions. In recent years, significant advancements in AI, machine learning, sensor fusion, and vehicle-to-everything (V2X) communication have propelled the development of fully autonomous vehicles. This paper explores the cutting-edge technologies driving the evolution of AVs, the challenges faced in their deployment, and the potential societal, economic, and regulatory impacts. It highlights the key innovations in perception systems, decision-making algorithms, and infrastructure integration, as well as the emerging trends towards Level 4 and Level 5 autonomy. The paper also discusses future directions, including ethical considerations and the roadmap to mass adoption of autonomous mobility. Ultimately, the integration of autonomous vehicles into global transportation systems is expected to revolutionize urban planning, reduce traffic accidents, and significantly lower emissions, paving the way for a smarter and more sustainable future.

**Introduction**

The evolution of autonomous vehicles (AVs) marks a monumental shift in the transportation sector, promising to transform how people and goods move across the globe. The advent of AVs, fueled by advancements in artificial intelligence (AI), machine learning (ML), sensor technology, and vehicle-to-everything (V2X) communication, has the potential to revolutionize mobility by enhancing safety, increasing efficiency, reducing environmental impact, and improving accessibility. As the global population grows, traffic congestion becomes a significant challenge, and the demand for sustainable transportation options intensifies, autonomous driving technology offers a compelling solution. Autonomous vehicles are defined as vehicles capable of operating without human intervention, with varying degrees of automation. According to the Society of Automotive Engineers (SAE) standard J3016, automation is classified into six levels, ranging from Level 0, where no automation exists, to Level 5, where full automation is achieved, and the vehicle operates independently in all conditions without human involvement. As AV technology progresses, most manufacturers and researchers focus on achieving higher levels of autonomy, particularly Levels 4 and 5, which promise a future where vehicles can operate in all environments, under any condition, without human control. The rise of AVs is not just a technological revolution but also a societal and economic one. Autonomous vehicles could lead to safer roads by significantly reducing human errors, which are responsible for more than 90% of traffic accidents. Moreover, AVs could potentially reduce congestion by optimizing traffic flow, decrease fuel consumption through more efficient driving patterns, and lower carbon emissions by integrating electric propulsion systems with smart driving algorithms. The automation of transportation could also redefine urban landscapes by enabling new forms of mobility and reshaping industries such as logistics, insurance, and public transport. Despite these promises, the path to full autonomy is fraught with challenges, including technical hurdles, ethical concerns, regulatory issues, and societal impact considerations. Autonomous vehicles rely on complex technologies, including machine learning algorithms for perception and decision-making, sensor fusion for environment mapping, and V2X communication for interacting with other vehicles and infrastructure. These technologies must be robust and reliable enough to handle the diverse and unpredictable environments encountered on the road. Moreover, AVs must be able to operate safely in all conditions, including inclement weather, complex traffic situations, and emergency scenarios, raising questions about the safety, reliability, and security of these systems. One of the most significant concerns is the ethical implications of autonomous driving. AVs must be programmed to make critical decisions in emergency situations, such as deciding how to avoid a collision when it is impossible to prevent harm to all parties involved. These decisions raise complex ethical questions about how AVs should be programmed to prioritize lives, mitigate harm, and conform to societal moral standards. This creates a need for clear ethical guidelines and regulatory frameworks to ensure that AV technologies align with public expectations and legal requirements. Furthermore, as AVs become increasingly connected and integrated into larger transportation ecosystems, cybersecurity becomes a critical concern. The potential for hacking, data breaches, and other malicious activities could undermine the safety and functionality of autonomous vehicles, leading to severe consequences. Ensuring the security of AV systems, both from external threats and from internal system failures, is therefore a priority for developers, regulators, and policymakers alike. The regulatory landscape for autonomous vehicles is still evolving, with various government agencies and organizations working to establish safety standards, guidelines, and laws governing the deployment of AVs. These standards need to balance innovation with safety and ensure that AVs are tested and validated in real-world conditions before they are allowed to operate at scale. However, the rapid pace of technological advancements in autonomous driving has often outpaced regulatory developments, leading to challenges in creating a cohesive and comprehensive regulatory framework.

This paper aims to provide an in-depth review of the current state of autonomous vehicle technology, focusing on the key areas of machine learning, sensor fusion, communication systems, ethical considerations, regulatory challenges, and safety concerns. The review will examine the advancements made in these areas, identify the challenges that remain, and explore the future directions of autonomous vehicle research and development. By synthesizing the current literature, this paper seeks to offer a comprehensive overview of the factors that will shape the future of autonomous vehicles, with a focus on their technological, societal, and regulatory implications. As autonomous vehicles continue to develop, it is essential for stakeholders in academia, industry, and government to collaborate in addressing the technological, ethical, and regulatory challenges associated with this transformative technology. The successful integration of AVs into society will require careful consideration of these factors, along with a commitment to ensuring that the benefits of autonomous driving technology are realized in a safe, secure, and equitable manner. This introduction sets the stage for an exploration of the key technologies and considerations surrounding autonomous vehicles, providing a framework for understanding the complex interplay of technological innovation, regulatory requirements, and societal impact that will define the future of transportation. As we move closer to a world with widespread AV adoption, it is critical to examine the progress made, the hurdles that remain, and the path forward to achieving the vision of fully autonomous, safe, and reliable transportation.

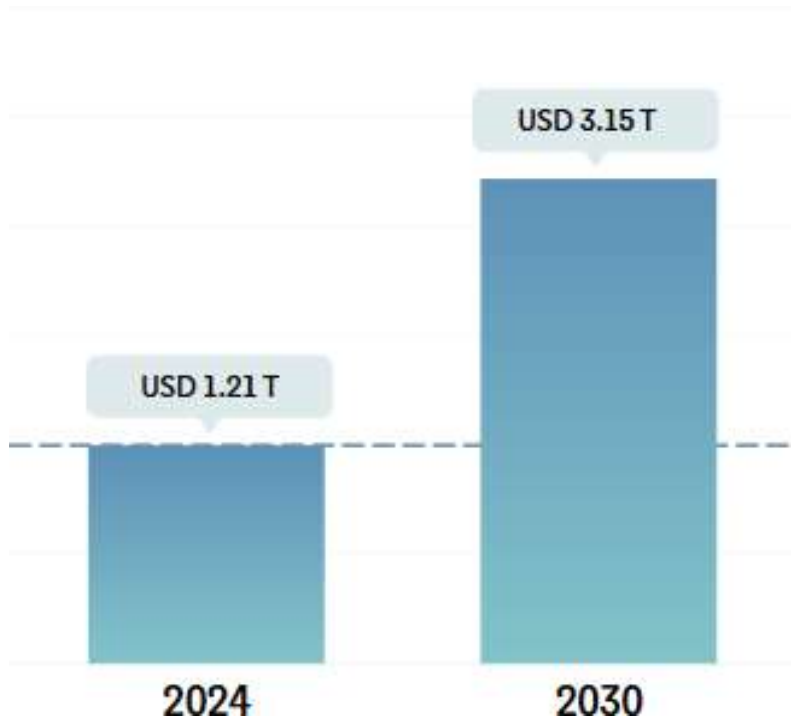

Fig.1: Market growth of Autonomous Vehicle Technology

**Literature Review**

Autonomous vehicles (AVs) are poised to revolutionize the transportation industry, promising a future where human intervention in driving is minimized or eliminated. The potential benefits include enhanced safety, increased efficiency, reduced environmental impact, and greater accessibility. However, achieving full autonomy in vehicles presents significant technological, ethical, regulatory, and societal challenges. This literature review explores the state of research in autonomous vehicle technologies, focusing on advancements in machine learning, sensor fusion, vehicular communication, ethics, and regulatory frameworks.

*1. Autonomous Vehicle Technologies and Machine Learning*

The development of autonomous vehicles heavily relies on advancements in artificial intelligence (AI), particularly machine learning (ML) and deep learning (DL). Chen and Zhang (2023) offer an extensive survey of the current state of AI in autonomous vehicles, detailing how deep learning models such as convolutional neural networks (CNNs) and recurrent neural networks (RNNs) are used for perception tasks, including object detection, path planning, and decision-making. These models enable vehicles to interpret complex environments in real time, a critical component for achieving safe and reliable autonomous navigation. Zhang and Zhou (2023) discuss the application of deep learning in the context of autonomous driving, highlighting challenges such as data scarcity, model generalization, and adversarial attacks. They emphasize the importance of robust training datasets and the continuous learning ability of models to adapt to new driving environments and conditions. Deep learning's role in AVs is not only in environmental perception but also in decision-making algorithms, which must handle uncertainty, predict human behavior, and make ethical decisions in complex traffic scenarios.

*2. Sensor Fusion in Autonomous Vehicles*

Sensor fusion is another cornerstone of AV technology, enabling vehicles to perceive their environment through a combination of sensors, including LiDAR, radar, cameras, and ultrasonic sensors. He and Zhang (2024) provide a detailed analysis of the challenges and future trends in sensor fusion for AVs. They discuss how combining data from various sensors helps overcome the limitations of individual sensor modalities. For instance, radar can perform well in adverse weather conditions, such as fog or rain, where cameras may fail. Conversely, cameras offer high-resolution visual information, which is essential for recognizing traffic signals, pedestrians, and other road users. Shladover (2023) provides an overview of the evolution of sensor technologies and sensor fusion techniques, arguing that achieving high levels of autonomy (Level 4 and 5) requires reliable, real-time integration of sensor data to handle complex driving environments. The integration of machine learning and sensor fusion is seen as a critical challenge in achieving the desired reliability and safety standards in autonomous driving systems.

*3. Communication Systems for Autonomous Vehicles*

The communication systems that support AVs are also undergoing significant development, with Vehicle-to-Everything (V2X) communication emerging as a key technology. Lee and Lee (2022) emphasize the importance of V2X communication for the safe operation of autonomous vehicles, where real-time data exchange between vehicles, infrastructure, and pedestrians helps improve situational awareness and coordination. V2X communication facilitates the exchange of information such as traffic signals, road conditions, and the positions of other vehicles, enabling AVs to make more informed decisions. Sun and Yang (2023) explore the potential of V2X communication systems in autonomous driving, discussing both the challenges and opportunities in terms of scalability, security, and network reliability. The authors highlight the need for robust communication protocols that can handle the massive data transfer required by autonomous vehicles, especially in densely populated urban environments.

*4. Ethical Considerations and Decision Making*

One of the most debated issues surrounding autonomous vehicles is the ethical dilemmas they may encounter while making decisions in emergency situations. Goodall (2023) addresses the need for a regulatory framework that guides AVs' decision-making in life-and-death scenarios. She argues that while automated systems can make decisions based on predefined ethical guidelines, there is still considerable debate about the moral principles that should guide these decisions. Autonomous vehicles must be able to navigate these challenges in a way that aligns with societal values and legal frameworks. Lin (2024) provides a comprehensive review of the ethical issues in autonomous vehicle design, discussing topics such as responsibility for accidents, privacy concerns, and the social implications of widespread AV adoption. The author emphasizes the need for international cooperation to develop consistent ethical standards and regulations to govern the deployment of autonomous vehicles.

*5. Regulatory and Legal Challenges*

The regulatory landscape for autonomous vehicles is still evolving, with various governments and organizations working to establish safety standards and guidelines for AV deployment. SAE International's J3016 standard (2024) provides a widely recognized taxonomy for autonomous driving systems, categorizing them into six levels of automation, from Level 0 (no automation) to Level 5 (full automation). This classification system has become the benchmark for defining the capabilities and limitations of AVs. The National Highway Traffic Safety Administration (NHTSA, 2023) has played a key role in developing safety standards for autonomous vehicles. In its recent report, NHTSA outlines the critical areas of focus for regulatory authorities, including vehicle safety, cybersecurity, and data privacy. Anderson and Kalra (2022) explore the challenges governments face in regulating autonomous vehicles, pointing out that the rapid pace of technological development often outstrips the ability of regulatory bodies to keep up. This creates a situation where AV manufacturers are often in a position to self-regulate, leading to concerns about safety, fairness, and accountability.

*6. Safety and Security Concerns*

Safety and security remain paramount concerns in the development and deployment of autonomous vehicles. Jafari and Wang (2024) offer a comprehensive study of the safety and security challenges faced by AVs, focusing on vulnerabilities to cyberattacks, sensor malfunctions, and system failures. They highlight the need for robust cybersecurity measures to protect AV systems from hacking and malicious interventions, particularly as vehicles become more connected and reliant on software-based control systems. The authors argue that ensuring the safety of autonomous vehicles will require multi-layered approaches that include secure communication protocols, regular software updates, and fail-safe mechanisms. The use of AI in critical driving functions also introduces new challenges in ensuring that these systems behave as expected under all conditions, including edge cases that may not be fully captured during testing.

*7. Impact on Society and Future Directions*

The widespread adoption of autonomous vehicles will have profound effects on various aspects of society, from urban planning and public transportation to insurance and employment. Shladover (2023) discusses the societal implications of AV deployment, emphasizing both the positive and negative consequences. On one hand, AVs could reduce traffic accidents, improve traffic flow, and reduce environmental impacts by optimizing fuel consumption. On the other hand, the transition to autonomous transportation could disrupt industries such as transportation, logistics, and insurance, leading to job displacement and economic upheaval. Rupp and Kühn (2022) outline potential future directions for autonomous vehicle technology, noting the importance of integrating AVs into existing transportation systems in a way that maximizes their benefits while minimizing negative consequences. This includes addressing challenges related to infrastructure development, data privacy, public acceptance, and the ethical deployment of AVs in various regions. Autonomous vehicles represent a significant technological advancement with the potential to transform the transportation sector. However, the road to full autonomy is fraught with challenges that span technological, ethical, regulatory, and societal domains. Advances in machine learning, sensor fusion, and communication systems are rapidly progressing, yet many obstacles remain in ensuring the safety, security, and reliability of autonomous vehicles. Moreover, ethical considerations, regulatory frameworks, and the societal impact of AVs require careful attention as we move toward a future where autonomous vehicles are a part of everyday life. Continued interdisciplinary research, collaboration between industry and regulators, and thoughtful consideration of the broader implications will be crucial for realizing the full potential of autonomous driving technology.

**Transition in Autonomous Vehicle Technology**

The transition to autonomous vehicle (AV) technology represents one of the most profound shifts in the history of transportation. It is not just a technological evolution but also a paradigm shift that challenges established notions of mobility, safety, urban design, and even societal norms. This transition is characterized by the gradual move from traditional vehicles driven by humans to fully autonomous vehicles that operate without human intervention. However, this shift is not a singular, sudden event but rather a phased transition that spans multiple stages of technological development, regulatory frameworks, societal readiness, and infrastructure adaptation. The development of autonomous vehicle technology has followed a gradual trajectory, with research and testing efforts spanning decades. This transition can be understood in terms of technological milestones, societal adoption, ethical concerns, and regulatory frameworks that must evolve alongside the vehicles themselves.

*1. Evolution of Autonomous Vehicle Technology*

A. Early Beginnings and Research

The concept of autonomous vehicles dates back to the mid-20th century, with the advent of early driver-assistance technologies such as cruise control and automated highway systems. However, the real breakthroughs in AV technology emerged in the 1980s and 1990s, with pioneering research in artificial intelligence (AI), machine learning (ML), and robotics. One of the earliest examples of autonomous driving was the "Navlab" project at Carnegie Mellon University, which tested self-driving cars in controlled environments. Similarly, the "Alvinn" project by the same university developed a self-driving vehicle capable of navigating a complex, cluttered environment. In the late 2000s, Google's Waymo project significantly accelerated the progress of autonomous vehicle research. In 2009, Google began testing self-driving cars equipped with laser radar (LiDAR), cameras, and other sensors, marking a pivotal moment in the transition towards fully autonomous vehicles. This development led to the successful demonstration of autonomous driving in real-world conditions, which set the stage for subsequent advancements.

B. Advancements in Machine Learning and Sensor Technology

A key driver of the transition in AV technology has been the integration of machine learning algorithms, which enable vehicles to process vast amounts of sensor data in real time and make decisions based on that information. Early autonomous vehicles relied heavily on pre-programmed decision trees and heuristic algorithms, which could only handle specific scenarios and required extensive mapping and controlled environments. However, with the advent of deep learning and computer vision technologies, AVs have become increasingly capable of handling dynamic, unpredictable environments. Deep learning, particularly convolutional neural networks (CNNs) and recurrent neural networks (RNNs), has played a crucial role in enhancing the perception capabilities of autonomous vehicles. These algorithms allow AVs to detect and classify objects, such as pedestrians, other vehicles, traffic signs, and road conditions, with a level of accuracy that surpasses human perception in certain contexts. Additionally, sensor fusion technologies, which combine data from various sensors like LiDAR, radar, and cameras, have become critical in enabling AVs to create a comprehensive,real-time model of their environment. This fusion of data provides AVs with the ability to make informed decisions, even in challenging conditions like poor weather or night driving.

C. From Testing to Deployment

The transition from testing to deployment has been one of the most challenging phases in the development of autonomous vehicles. Early autonomous vehicle projects were limited to closed environments or specific, low-speed scenarios, such as autonomous shuttles in controlled areas or autonomous trucks in well-marked, predefined routes. However, recent advancements have pushed AVs into more complex environments, with companies like Waymo, Tesla, and Cruise deploying autonomous vehicles on public roads in various cities. Despite these advancements, fully autonomous vehicles are still far from ubiquitous on the roads. Current AVs typically operate under the "Level 4" automation (highly automated driving) in specific geofenced areas, with limited intervention required from human drivers. Tesla's Full Self-Driving (FSD) system, for example, operates under "Level 2" or "Level 3" automation (partial to conditional automation), where the vehicle can drive itself in specific conditions, but the human driver must remain alert and ready to take control. The transition to "Level 5" autonomy, where vehicles can operate in any environment and under any condition without human intervention, remains a goal that is still being actively researched and developed. This final stage of transition presents the most significant technological challenges, as AVs must be capable of understanding and responding to a virtually infinite number of potential road scenarios, all while ensuring safety, security, and reliability.

2. *Societal and Economic Transition*

A. Changing Human Roles and Perceptions

The transition to autonomous vehicles brings with it profound societal and economic implications. One of the most significant changes is the role of humans in transportation. In the traditional model, humans act as both drivers and decision-makers in real-time traffic situations. However, with the advent of AVs, this role is being supplanted by advanced algorithms and AI. This shift could lead to a reduction in traffic accidents caused by human error, which is responsible for more than 90% of road crashes. However, the loss of the driving role raises concerns about job displacement, especially in sectors such as trucking, ride-hailing services, and delivery logistics. According to estimates, millions of driving-related jobs could be impacted by the widespread adoption of autonomous vehicles. As a result, the transition to AVs will require a broader societal discussion on how to manage the economic displacement of workers and the redistribution of employment opportunities. Public policy, including retraining programs and labor protections, will be essential to mitigate the negative social impacts of this transition. Moreover, the shift to AVs could alter societal attitudes towards mobility. For instance, as autonomous vehicles reduce the burden of driving, people may see a shift in their relationship with transportation, potentially leading to an increase in shared mobility services and a decrease in personal vehicle ownership. This change could have wide-reaching implications for urban planning, with a shift away from car-centric infrastructure to more walkable, bike-friendly, and transit-oriented cities.

B. Accessibility and Equity Considerations

Autonomous vehicles also have the potential to improve mobility for underserved populations, including the elderly, disabled, and people in rural areas. For these groups, the availability of autonomous vehicles could provide a new level of independence and freedom, reducing their reliance on public transportation or human-driven taxis. However, ensuring equitable access to these technologies is a challenge that needs to be addressed during the transition period. For instance, there are concerns about the affordability of AVs and the potential for social inequality if the benefits of this technology are not distributed equitably across all demographics.

C. Infrastructure Adaptation

The successful deployment of autonomous vehicles also depends on the adaptation of urban infrastructure. AVs require advanced communication systems to interact with their environment, such as Vehicle-to-Everything (V2X) communication, which allows vehicles to communicate with traffic signals, road signs, and other vehicles. In addition, roads, bridges, and traffic management systems need to be updated to accommodate the unique needs of AVs, such as lane markings, road sensors, and high-speed data networks. This adaptation requires significant investment from both public and private sectors.

*3. Regulatory and Ethical Challenges*

A. Legal and Regulatory Frameworks

The transition to autonomous vehicle technology is also heavily influenced by the legal and regulatory environment. Regulatory bodies such as the National Highway Traffic Safety Administration (NHTSA) and the Society of Automotive Engineers (SAE) have developed frameworks to define the different levels of vehicle automation, as well as safety standards. However, the rapid pace of technological advancement often outstrips regulatory developments, creating challenges for lawmakers and safety regulators in ensuring that AVs meet appropriate safety standards. One key challenge in this area is the creation of universal, international standards for autonomous driving technology. Because AVs will likely operate across various jurisdictions with different traffic laws, regulatory frameworks must be harmonized to ensure safe and efficient global deployment. Additionally, AV manufacturers must navigate liability issues, such as who is responsible in the event of an accident involving an autonomous vehicle. These legal and ethical concerns are integral to the smooth transition of AVs from research and testing to mainstream adoption.

B. Ethical Dilemmas

Ethical dilemmas also play a central role in the transition to AVs. One of the most significant ethical challenges involves programming AVs to make decisions in emergency situations, such as choosing between two unavoidable collisions. The ethical implications of these decisions are complex and multifaceted, requiring careful consideration of societal values, legal standards, and individual rights. Researchers and policymakers must address how AVs should be programmed to ensure that they make morally and legally acceptable decisions in life-and-death situations.

Conclusion

The transition to autonomous vehicle technology is a multi-dimensional process that involves technological, societal, economic, regulatory, and ethical challenges. While AVs promise significant benefits in terms of safety, efficiency, and accessibility, their successful integration into society requires careful planning, collaboration, and forward-thinking strategies. As the technology continues to evolve, it is crucial for stakeholders to work together to ensure that the transition to fully autonomous vehicles is smooth, equitable, and beneficial for all members of society. The future of transportation will depend not only on technological advancements but also on how well society adapts to and embraces these changes.

**Real-Life Case Studies on Autonomous Vehicle Technology** In the real-world development of autonomous vehicles (AVs), several companies, cities, and countries have conducted various pilot projects and testing programs. These case studies provide valuable insights into the challenges and opportunities of transitioning to autonomous driving technology. The following table highlights key examples of real-life AV projects, examining the scope of each initiative, the technologies involved, the outcomes, and the lessons learned.

| Case Study | Location | Company | Project Overview | Technologies | Key Outcomes | Challenges | Lessons Learned |
|---|---|---|---|---|---|---|---|
| **Waymo Self-Driving Cars** | Phoenix, AZ, USA | Waymo (Google) | Waymo operates autonomous taxis with no safety driver. | LiDAR, cameras, AI, sensor fusion | Over 20 million miles driven with low incident rate. | Safety concerns, public trust, limited testing areas. | Real-world testing builds confidence, public outreach is key. |
| **Tesla Autopilot and FSD** | Nationwide, USA | Tesla | Tesla's Autopilot provides partial autonomy, Full Self-Driving is in development. | Cameras, radar, AI, machine learning | Over 5 billion miles driven in Autopilot mode. | Still needs human oversight, public skepticism. | Iterative updates and human oversight are essential. |
| **Cruise Autonomous Ridesharing** | San Francisco, CA, USA | Cruise (GM) | Testing autonomous cars for ridesharing in urban areas. | LiDAR, cameras, AI, deep learning | Permit for autonomous ridesharing, expanding coverage. | Complex urban environments, regulatory approval. | Collaboration with regulators is crucial for scaling. |
| **Navya Autonomous Shuttles** | France, USA, Japan | Navya | Low-speed autonomous shuttles for public transport. | GPS, LiDAR, cameras, AI | Successful integration into public transport networks. | Low speed, navigating mixed traffic. | Gradual deployment in controlled environments works. |
| **Baidu Apollo Pilot Program** | Beijing, Changsha, China | Baidu | Self-driving taxis for ride-hailing in select cities. | LiDAR, AI, V2X, HD mapping | Expanding to more cities, driverless taxis in operation. | Regulatory hurdles, complex road conditions. | Government support and clear regulations are key. |
| **Waymo-Jaguar I-PACE Partnership** | Phoenix, AZ, USA | Waymo, Jaguar Land Rover | Partnership to integrate autonomous tech in Jaguar I-PACE EVs. | LiDAR, radar, AI, machine learning | Fleet expansion with electric autonomous vehicles. | Integration challenges, balancing safety and performance. | EVs align well with autonomy, partnerships accelerate deployment. |
| **Uber ATG Delivery** | Pittsburgh, PA, USA | Uber ATG (Aurora) | Autonomous trucks for freight delivery. | LiDAR, cameras, AI, GPS | Pilots in freight transport, focusing on long-haul. | Safety concerns, regulatory delays. | Autonomous freight could disrupt logistics. |

*Analysis of Real-World Case Studies*

1. **Technological Integration and Advancements**: Across all case studies, the integration of various sensors—such as LiDAR, radar, and cameras—has been critical for AV functionality. Most projects rely heavily on sensor fusion, AI, and machine learning algorithms to ensure reliable object detection, path planning, and decision-

   making. Additionally, the rise of 5G and V2X communication is enhancing the ability of AVs to interact with surrounding infrastructure, including traffic lights, road signs, and other vehicles.
2. **Challenges in Public Perception and Trust**: One consistent challenge in the adoption of AVs is public trust. Cases like Tesla's Autopilot and Waymo's ride-hailing service show that, while AVs can handle certain tasks with remarkable efficiency, human oversight remains necessary. Regulatory bodies and companies are actively working to address concerns around safety and liability.
3. **Regulatory and Legal Frameworks**: Different jurisdictions have varied regulatory approaches. In the United States, regulatory approval for AVs, especially for driverless operations, has been slow, as seen in the cases of Uber and Cruise. In China, government support has accelerated the adoption of AVs, as evidenced by Baidu's Apollo Go service.
4. **Economic Impacts**: The automation of transportation, particularly in sectors like freight, has the potential to disrupt jobs and industries. While AVs could create new jobs in tech development and vehicle manufacturing, jobs in trucking and delivery services are likely to be displaced, as evidenced in Uber Freight's pilot with autonomous trucks.

These real-life case studies show that while autonomous vehicle technology has made significant progress, widespread adoption still faces challenges. Technological hurdles, regulatory uncertainties, public perception, and integration with existing infrastructure remain key obstacles. However, each case study offers valuable lessons on how to overcome these challenges, with particular emphasis on safety, stakeholder collaboration, and gradual deployment. As the technology matures and regulatory frameworks evolve, the successful integration of autonomous vehicles into mainstream transportation systems appears increasingly feasible.

**Opportunities, Challenges, and Future Scope of Autonomous Vehicle Technology**

The development of autonomous vehicle (AV) technology presents a transformative opportunity across multiple sectors, offering substantial benefits to transportation, mobility, and society at large. However, the transition to widespread adoption is also accompanied by numerous challenges, ranging from technical and regulatory obstacles to societal and ethical concerns. The future scope of autonomous vehicles is vast, but their success will depend on overcoming these challenges and tapping into the opportunities they present.

*Opportunities*

1. **Enhanced Road Safety and Reduced Traffic Accidents**
    - **Opportunity**: One of the most compelling advantages of autonomous vehicles is the potential for improving road safety. Human error is responsible for the vast majority of traffic accidents, including collisions caused by distraction, impaired driving, fatigue, and poor decision-making. Autonomous vehicles, which rely on advanced sensors, machine learning, and AI to navigate, can significantly reduce the likelihood of these accidents.
    - **Impact**: According to estimates from the National Highway Traffic Safety Administration (NHTSA), human error accounts for over 90% of crashes. By automating the driving process, AVs could prevent thousands of deaths and injuries each year, improving overall traffic safety.

*Narisetty Revolutionizing Mobility: The Latest….*Venkata Sai Chandra Prasanth 1364- o **Example**: Waymo's autonomous taxis in Phoenix, Arizona, have been operating without a safety driver in certain areas, achieving millions of miles driven with a lower rate of incidents compared to human-driven vehicles.

2. **Reduced Traffic Congestion**
   - o **Opportunity**: Autonomous vehicles can optimize traffic flow by communicating with each other and traffic infrastructure through Vehicle-to-Everything (V2X) communication. This inter-vehicle communication allows AVs to anticipate traffic patterns, adjust speeds, and merge more efficiently, reducing congestion and improving traffic flow.
   - o **Impact**: In a world of increasing urbanization, reducing traffic congestion could lead to significant time savings, increased productivity, and less stress for commuters. AVs also promise to reduce fuel consumption and lower carbon emissions by optimizing routes and driving patterns.
   - o **Example**: Studies have shown that autonomous vehicles could reduce traffic congestion by up to 30%, as they would drive more efficiently, reducing bottlenecks and delays.

3. **Increased Mobility for the Elderly and Disabled**
   - o **Opportunity**: Autonomous vehicles hold the potential to transform mobility for populations that are currently underserved by traditional transportation systems, such as the elderly, disabled, and those with limited access to public transit. AVs could provide greater independence for people who are unable to drive due to age, disability, or other factors.
   - o **Impact**: The availability of autonomous ride-hailing services could eliminate the need for caregivers to provide transportation and improve the quality of life for individuals who are otherwise isolated.
   - o **Example**: In pilot programs like those run by Waymo, elderly and disabled users have been able to access autonomous taxis, providing them with greater autonomy in their daily activities.

4. **Environmental Benefits**
   - o **Opportunity**: Autonomous vehicles, particularly when paired with electric powertrains, have the potential to drastically reduce emissions and environmental pollution. By optimizing routes and driving behavior, AVs can reduce fuel consumption, leading to fewer carbon emissions. Additionally, the widespread adoption of electric autonomous vehicles (EVs) could help decarbonize the transportation sector.
   - o **Impact**: The integration of autonomous EVs into urban mobility systems could help cities achieve sustainability goals and contribute to global efforts to mitigate climate change.
   - o **Example**: Waymo's partnership with Jaguar Land Rover to develop a fleet of autonomous electric vehicles (EVs) is an example of how AV technology can contribute to a more sustainable transportation ecosystem.

5. **New Business Models and Economic Opportunities**
   - o **Opportunity**: The rise of autonomous vehicles will foster new business models and economic opportunities across industries. This includes new forms of mobility services such as autonomous ride-hailing, shared mobility, and even last-mile delivery solutions. Additionally, autonomous vehicles could create new markets in AV-related industries like vehicle manufacturing, software development, and cybersecurity.
   - o **Impact**: The autonomous vehicle market could generate significant economic value by disrupting industries like logistics, transportation, insurance, and retail. New

*Nanotechnology Perceptions* **20 No. S12**(2024) 1354–1367

- opportunities for innovation in AI, sensor technology, and communication systems will also emerge.
- **Example**: Companies like Uber, Lyft, and Cruise are already investing in autonomous ridesharing and delivery services, reshaping how goods and people move within cities.

*Challenges*

1. **Technical and Engineering Challenges**
   - **Challenge**: While autonomous vehicles have made significant advancements, there are still major technical hurdles to overcome. AVs need to be capable of navigating complex, dynamic environments—such as urban streets, inclement weather, and unpredictable human behavior—with the same level of reliability as human drivers.
   - **Key Issues**: Challenges include perfecting object recognition, improving sensor fusion (e.g., combining data from cameras, radar, and LiDAR), developing robust decision-making algorithms for complex traffic scenarios, and ensuring the vehicle can safely interact with human drivers, cyclists, and pedestrians.
   - **Example**: Tesla's Full Self-Driving (FSD) system, while promising, still requires human supervision due to its limitations in handling more complex scenarios (e.g., recognizing unusual road signs or navigating in heavy traffic).
2. **Safety, Security, and Cybersecurity Concerns**
   - **Challenge**: The safety of autonomous vehicles is a paramount concern. While AVs can significantly reduce human errors, the vehicles themselves must be free of vulnerabilities that could lead to accidents or malfunctions. Cybersecurity risks are particularly relevant, as AVs rely on complex software and communication networks that could be targeted by hackers.
   - **Key Issues**: Ensuring the integrity of AV systems against cyberattacks, safeguarding data privacy, and addressing vulnerabilities that could allow malicious actors to take control of the vehicle are critical concerns.
   - **Example**: In 2015, researchers demonstrated the potential for remote hacking of a Jeep Cherokee, highlighting the risks associated with connected, software-dependent vehicles.
3. **Regulatory and Legal Challenges**
   - **Challenge**: The regulatory landscape for autonomous vehicles is still evolving. Governments and regulatory bodies must create clear frameworks for testing, deploying, and monitoring AVs on public roads. These regulations will need to balance safety, innovation, and public concerns while addressing issues such as liability in the event of an accident.
   - **Key Issues**: Lack of standardized regulations across jurisdictions, uncertainty regarding liability in accidents, and defining the appropriate levels of automation for different use cases remain significant obstacles to widespread deployment.
   - **Example**: In the United States, the National Highway Traffic Safety Administration (NHTSA) has been slow to release comprehensive guidelines, and states like California have struggled to define clear rules for autonomous vehicle testing.
4. **Public Trust and Acceptance**
   - **Challenge**: Public perception of autonomous vehicles is mixed, with concerns over safety, reliability, and the ethics of AV decision-making in emergency situations. Building trust in the technology is crucial for widespread adoption.
   - **Key Issues**: Ensuring that AVs are perceived as safe, ethical, and trustworthy will require transparency in testing, regulatory oversight, and clear communication with the public.

- o **Example**: High-profile accidents involving semi-autonomous systems, such as the fatal Uber self-driving car crash in 2018, have fueled skepticism about the readiness of AVs for widespread deployment.
- o **Infrastructure and Urban Planning Challenge**: The successful integration of autonomous vehicles requires substantial changes in road infrastructure, traffic management systems, and urban planning. AVs rely on accurate mapping and real-time data from infrastructure elements such as traffic lights, road signs, and other vehicles.
- o **Key Issues**: Cities must adapt their infrastructure to accommodate AVs, including developing smart roads, integrating V2X communication, and ensuring that AVs can function safely in environments with mixed traffic.
- o **Example**: In cities like San Francisco, where autonomous vehicles are being tested, the lack of standardized road markings, unclear signage, and unpredictable human drivers pose challenges for AV systems.

*Future Scope*

1. **Level 5 Autonomy (Full Autonomy)**
    - o **Scope**: Achieving Level 5 autonomy, where vehicles can operate without any human intervention in all environments and conditions, remains the ultimate goal of AV research. This will require advances in AI, machine learning, and sensor technologies to handle complex, dynamic, and unpredictable road environments.
    - o **Impact**: Level 5 vehicles could operate in fully driverless fleets, providing seamless, on-demand transportation services that transform urban mobility.
2. **Integration with Smart Cities**
    - o **Scope**: Autonomous vehicles will play a key role in the development of smart cities. AVs will integrate with IoT (Internet of Things) systems and V2X technologies to create more efficient, sustainable, and connected urban environments. Cities will use data from AVs to optimize traffic flow, reduce congestion, and improve public services.
    - o **Impact**: The development of smart cities could significantly improve the efficiency of urban mobility systems, reduce energy consumption, and enable a higher quality of life for residents.
3. **Autonomous Freight and Delivery**
    - o **Scope**: Autonomous trucks and delivery vehicles offer immense potential for transforming logistics and supply chains. Long-haul trucking, which accounts for a large portion of carbon emissions in the transportation sector, could be revolutionized by autonomous trucks that reduce fuel consumption, increase delivery efficiency, and reduce labor costs.
    - o **Impact**: Automation in freight transport could lead to cost savings, more efficient delivery networks, and potentially faster deliveries, which could benefit industries like e-commerce and manufacturing.
4. **Ethical AI and Governance**
    - o **Scope**: As autonomous vehicles become more integrated into society, the ethical implications of AI-driven decision-making must be carefully considered. Researchers are working on frameworks for ensuring that AVs make decisions that align with societal values, particularly in emergency situations (e.g., how should an AV choose between two equally harmful collision scenarios?).
    - o **Impact**: Developing ethical frameworks and governance structures for AVs will help ensure that autonomous systems act in ways that align with public safety and fairness.

## Conclusion

Autonomous vehicle (AV) technology holds significant promise for transforming transportation by improving safety, reducing congestion, increasing mobility for underserved populations, and contributing to environmental sustainability. However, its widespread adoption faces challenges, including technical limitations, safety concerns, regulatory hurdles, and public trust issues. Real-world case studies, such as Waymo, Tesla, Cruise, and others, demonstrate both the potential and the obstacles of AVs. These initiatives highlight the importance of rigorous testing, clear regulatory frameworks, and public engagement. As technology advances and challenges are addressed, autonomous vehicles are likely to play a key role in reshaping the future of transportation, driving innovation across industries and society.